\documentclass[]{article}

\pdfoutput=1
 
\usepackage{color}
\usepackage[english]{babel}
\usepackage{amsmath}
\usepackage{amssymb}
\usepackage{amsfonts}
\usepackage{graphicx}
\usepackage[nodots]{numcompress}
\usepackage{multirow}

\begin{document}


\title{Reliability Analysis of Processes with Moving
Cracked Material}

\author
{\small M. Tirronen\\
\small University of Jyv\"askyl\"a,\\
\small  Department of Mathematical Information 
\small Technology,\\
\small P.O. Box 35 (Agora),
\small FI--40014 University of Jyv\"askyl\"a,
\small Finland\\
\small maria.j.e.tirronen@jyu.fi}

\maketitle

\begin{abstract}
The reliability of processes with moving elastic and isotropic 
material containing initial cracks is considered in terms of fracture. 
The material is modelled as a moving plate which is simply supported from two of its sides and
 subjected to homogeneous tension acting in the travelling direction.
For tension, two models are studied: i) tension is constant with respect to time, and ii) 
tension varies temporally
 according to an Ornstein-Uhlenbeck process. {\color{black} Cracks of random length are assumed} to
 occur in the material according to a stochastic counting process. 
For a general counting process, a representation of the nonfracture probability of the system is obtained that exploits conditional Monte Carlo simulation. Explicit formulae are derived for special cases.
 To study 
 the reliability of the system with temporally varying tension, a 
 known explicit result for the first 
 passage time of an Ornstein-Uhlenbeck 
     process to a constant boundary is utilized. Numerical examples are provided for 
      printing presses
     and paper material.
\end{abstract}



\section{Introduction}

There are systems in industry in which material moves unsupportedly 
between two rollers under a longitudinal edge tension. {\color{black} 
Such systems can be found, e.g., in manufacturing and printing of paper. In paper machines and printing presses, the tension is essential for the transport of the material and it is created by a velocity difference of the rollers. The relative velocity difference of the rollers is called draw, and the span between the rollers is called an open draw.}

{\color{black} To achieve good productivity in systems with moving material, there is a 
demand for running the system at a high speed but at the same time avoiding 
runnability problems.
In pressrooms, 
runnability
problems include web breaks, register errors, wrinkling 
and the instability of the paper web \cite{ParolaEtAl2000}. Of these problems, especially web breaks have gained attention in the print industry \cite{Uesaka2013}.

One of the suspected causes of web breaks in pressrooms are defects.
Defects 
in a paper web 
can be classified into two categories: microscopic and macroscopic defects. Microscopic defects originate from the natural disorder in paper, such as
formation, local fibre orientation and variation of wood species \cite{Salminen2010}. Macroscopic defects are introduced during the papermaking and transportation processes. In papermaking, a condensation drip in pressing or drying section or a lump on press rolls or press felt can cause holes in the paper web \cite{Smith1995book}. Such defects occur randomly or in a fixed pattern. Stress formed from running a high roll edge through a nip may cause cracks on the edge of the paper web \cite{Smith1995book}. Edge cracks of such origin typically occur randomly in the sheet. Insufficient roll edge protection during handling and storage may also cause edge cracks. A cut or nick in the edge of the roll cause multiple edge cracks in the sheet in a localized area \cite{Smith1995book}. 

Web breaks occur at random intervals and they are rare events in pressrooms \cite{PageSeth1982}. Thus, data from a large number 
of rolls is required for determining the causes of web breaks with a reasonable level of confidence \cite{DengEtAl2007} and such data is difficult to obtain under controlled conditions \cite{UesakaFerahi1999}. In addition to the rarity of web breaks, there are often many dependent random variables involved in the printing process, and controlling of them may appear difficult \cite{UesakaFerahi1999}. To avoid these problems, two approaches for finding causes of web breaks have been suggested \cite{UesakaFerahi1999}. One is to conduct data-analysis on massive pressroom databases and the other is to investigate the effect of different factors on web breaks by mathematical modelling.

Although the effect of macroscopic defects have gained attention in the research (see, e.g., literature review in \cite{Uesaka2013}), to the author's knowledge, only a few studies aim to predict the connection of macroscopic defects and web breaks by mathematical modelling. Swinehart and Broek \cite{swineharBroekt1996} developed a web break model, based on fracture mechanics, which included the size distribution of flaws, web strength and web tension. In \cite{swineharBroekt1996}, the tension was regarded as constant. Uesaka et al. \cite{UesakaFerahi1999} studied the effect of cracks on web breaks by a break-rate model based on the weakest link
theory of fracture. The number of breaks per one roll during a run was derived by considering the strength of charateristic elements of the web. In \cite{UesakaFerahi1999} the tension in the system was assumed to be constant and later, Hristopulos and Uesaka \cite{HristopulosUesaka2002} presented a dynamic model of the web
transport derived from fundamental physical laws. In
conjunction with the weakest link fracture
model, the model allows investigating the impact of tension variations on web break rates.
 
The break-rate model used in \cite{UesakaFerahi1999,HristopulosUesaka2002} predicts the upper estimate of the break frequency.
However, considering an upper bound of fracture probability may lead to an overconservative upper bound for a safe range of tension. The studies of mechanical instability suggest that the higher the tension, the higher the velocity of the moving material can be \cite{BanichukEtAl2009-IJSS}. Thus, from the view point of maximal production, an overconservative tension is undesirable as it underestimates the maximal safe velocity.

Motivated by paper industry, defects have also gained attention in the studies of instability of moving materials. Banichuk et al. \cite{BanichukEtAl2012-MBDSM} studied an elastic and isotropic plate that has initial cracks of bounded length travelling in a system of rollers. In \cite{BanichukEtAl2012-MBDSM}, the plate was assumed to be subjected to constant or (temporally) cyclic in-plane tension and the Paris' law was used to describe the crack growth induced by tension variations. The optimal average tension was sought for the maximum crack length by considering a productivity function which takes into account both instability and fracture. 
Moreover, an attempt to take the stochasticity of systems with moving material into account was made in the study by Tirronen et al. \cite{TirronenEtAl2014} in which the safe 
transition of elastic and isotropic material with initial cracks was analyzed by modelling the problem parameters as random variables. In \cite{TirronenEtAl2014}, critical regimes for 
the tension and velocity of the material were sought by considering the probabilities of fracture and instability.

Although tension in a printing press is known to change in time due to draw variations 
\cite{Uesaka2013} and tension fluctuations have been suggested to cause web breaks \cite{Uesaka2004}, the tension in the system was regarded as constant in \cite{UesakaFerahi1999,swineharBroekt1996}. In \cite{TirronenEtAl2014}, the tension was assumed to be constant while a crack travels through an open draw although the constant value was assumed to include uncertainty. In \cite{BanichukEtAl2012-MBDSM}, only deterministic variations of tension were considered although the draw variations contain white noise in addition to specific high/low frequency components \cite{Uesaka2013}. In a printing press, cyclical tension variations may be caused by out-of-round unwind rolls or vibrating machine elements such as unwind stands (see \cite{roisum1990} and the references therein). In addition to cyclical variations, tension may vary aperiodically due to poorly tuned tension controllers, drives, or unwind brakes (\cite{roisum1990} and the references therein). The net effect of such factors cause the tension to fluctuate around the mean value \cite{roisum1990}.

This study aims at developing mathematical models for systems in which a moving cracked material travels under longitudinal tension. The material is assumed to be elastic and isotropic, and the models of this study focus on describing the occurrence of defects in the material and tension variations in the system, taking into account the stochasticity of these phenomena.
This paper extends the study \cite{TirronenEtAl2014} 
by modelling the crack occurrence and temporal variations of tension by stochastic processes, which 
enables examination of system longevity. Instead of estimating the fracture probability from above, the present paper aims at directly computing the fracture probability predicted by the model.

Two different models 
are considered for temporal value of tension. The first 
model describes tension as constant with respect to time. The second model describes the tension as a stationary 
Ornstein-Uhlenbeck process. With the latter model, tension has a constant mean value, the set tension, 
around which it fluctuates temporally. The Ornstein-Uhlenbeck process 
can be considered as the continuous-time 
analogue of the discrete-time AR(n) process.
 It provides a mathematically well-defined continuous-time model for fluctuations of systems whose measurements contain  white noise \cite[Chapter 4]{gardiner1983}. Moreover, a stationary process describes random fluctuations of a system which has settled down to a steady state and whose statistical properties do not depend on time \cite[Sections 3.7]{gardiner1983}.
The stationary Ornstein-Uhlenbeck process can be regarded as a simplified model of tension variations in a printing press.

In this study, we consider straight-line through-thickness cracks perpendicular to the travelling direction and located on the edge of the material. 
Sharp edge cracks oriented in the cross direction of the paper web are most critical in printing presses \cite{roisum1990b}.
Other stochastic quantities in the presented model describe the occurence of cracks in the open draw and the lengths of the cracks. The locations of the cracks in the travelling direction are described by 
a stochastic counting process.
The lengths of the cracks are modelled
by independent and identically distributed (i.i.d.) random variables.}

The reliability of the system is studied in terms of fracture by applying
  linear elastic fracture mechanics (LEFM). For a general counting process,
  the nonfracture probability is obtained 
by utilizing conditional Monte Carlo simulation which is one of the most effective techniques for variance reduction \cite[Section 5]{RubinsteinKroesebook2007}. An explicit representation is {\color{black}derived} for a few 
special cases. When 
there is stochastic volatility in tension,  
considering the probability of a fracture leads to first passage time problems which are solved by exploiting the spectral expansion of the first hitting time of an Ornstein-Uhlenbeck 
process to a constant boundary, as given in \cite{Linetsky2004}.

{\color{black}Numerical examples are computed for material and machine parameters typical of dry paper (newsprint) and printing presses. The reliability of the system is studied with different models for crack occurrence. The impact of different parameters of the stochastic quantities on the reliability of the system is illustrated.
} 

\section{Problem setup}

In this study, we consider a moving elastic and isotropic band containing initial cracks
during its transition
through an open draw. Below, a mathematical model for the moving band 
is presented.
The model is 
similar to the one presented in \cite{BanichukEtAl2009-IJSS}.

\begin{figure}[h!]
\includegraphics[width=0.6\textwidth]{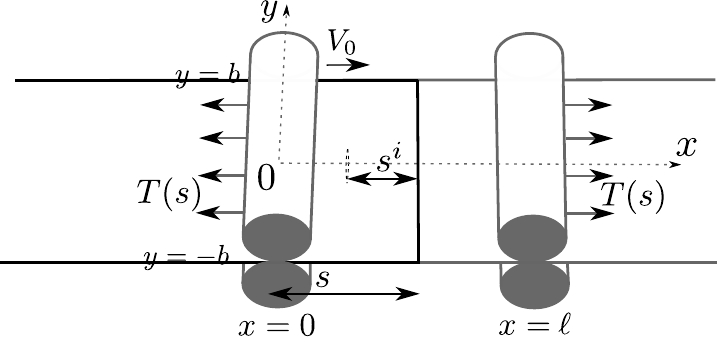}
%
%
\caption{The part of the band that is travelling in the open draw is modelled as
a plate tensioned at the supported edges with the homogeneous tension
$T(s)$. The minimum distance between the $i$th crack and the first end of the band is denoted by $s^i$. 
The drawing is adapted from Fig. $1$ in \cite{TirronenEtAl2014}.
}
\label{initials:fig1}       
\end{figure}

{\color{black}To study the behavior of the band in the open draw, consider a rectangular
part of it that occurs between the supports momentarily:}
\begin{equation}
\mathcal{D}=\{(x,y):\hspace{0.5em}0<x<\ell,\hspace{0.5em}-b<y<b\}\label{plate-element}
\end{equation}
in $x,y$ coordinates, see Fig. \ref{initials:fig1}. The length of the span between the supports is $\ell$ and the width of the band is $2b$. 
The part $\mathcal{D}$ is modelled as an elastic and isotropic plate that has constant thickness $h$
 and Young modulus $E$.
The sides of the plate
\begin{equation}
\{x=0,\hspace{0.5em} -b< y < b\}\hspace{0.5em}\text{and}\hspace{0.5em}\{x=\ell,\hspace{0.5em}-b< y< b\}\label{supported-boundaries}
\end{equation} 
are simply supported, and the sides 
\begin{equation}
\{y=-b,\hspace{0.5em} 0< x< \ell\}\hspace{0.5em}\text{and}\hspace{0.5em}\{y=b,\hspace{0.5em}0< x< \ell\}
\end{equation}
are free of tractions.

{\color{black}\subsection{Tension}

The plate element \eqref{plate-element}} is subjected to homogeneous tension acting in the $x$ direction. Two different models 
describing the temporal value of tension are studied. In the first model, the value of 
tension is assumed to be a constant $T_0>0$. 
In the second model,
the tension exhibits temporal random fluctuations. In this case, the tension is described by a continous-time stochastic process
\begin{equation}
T=\{T(s),\hspace{0.5em}s\ge 0\}\label{tension-process}
\end{equation}
in a probability space $(\Omega,\mathcal{F},\mathbb{P})$. Above, $s$ denotes the length of 
the part that has travelled through the 
first end of the open draw, see Fig. \ref{initials:fig1}. 

{\color{black}Furthermore, temporal random fluctuations of tension are described by a stationary Gaussian Markov process. 
A stationary process describes the stochastic fluctuations of a system which has settled down to a steady state and whose statistical properties do not depend on time \cite[Section 3.7]{gardiner1983}.  Gaussian random variables approximate many real-life variables adequately due to the central limit theorem \cite[Section 2.8.2]{gardiner1983}. Moreover, Markov processes can be used to describe many real systems which have small memory times (see \cite[Sections 3.2 and 3.3]{gardiner1983}). 

With these assumptions, a natural model for the tension is a stationary Ornstein-Uhlenbeck process. The stationary Ornstein-Uhlenbeck process is the only one-dimensional stochastic process that is stationary, Gaussian and Markovian \cite[Section 3.8.4]{gardiner1983}.
With the Ornstein-Uhlenbeck process, the tension changes with respect to $s$ according to the stochastic differential equation
\begin{equation}
dT(s)=a_T(T_0-T(s))ds+\sigma_T dW(s),\label{ou}
\end{equation}
where $W$ is the standard Brownian motion (Wiener process) and $T_0$, $a_T$ and $\sigma_T$  are strictly positive constants. The parameter $T_0$ is the long-term mean of the process, the coefficient $a_T$ is the rate by which the process $T$ reverts 
toward $T_0$ and $\sigma_T$ describes the degree of volatility around $T_0$. In the following, the long-term mean $T_0$ is also called the set tension. Furthermore, the process $T$ is stationary
if the initial value satisfies
\begin{equation}
T(0)\sim\mathcal{N}\bigg({T_0,\frac{\sigma_T^2}{2a_T}}\bigg),\label{theta-initial}
\end{equation}
where $\mathcal{N}$ is the normal distribution \cite[Section 3.3.1]{Glasserman2003}.

Since $T$ is stationary, the probability density function of $T(s)$ is time-independent. We denote the probability density function of $T$ by $f_T$. By denoting the coefficient of variation of $T(s)$ (the mean of $T(s)$ divided by its standard deviation) by $c_T$, we have
\begin{equation}
\frac{\sigma_T}{\sqrt{2a_T}}=c_TT_0.\label{c-T}
\end{equation}

The transition probability density of
$T$ (the conditional density of $T(t+s)$ given $T(s)=x$) is given by the formula
\begin{align}
p(t,x,y)=&\frac{1}{\sqrt{\pi\sigma_T^2\big(1-\exp[-2a_Tt]\big)/a_T}}\cdot\nonumber\\
&\cdot\exp\bigg[-\frac{\big(y-T_0-(x-T_0)\exp[-a_Tt]\big)^2}{\sigma_T^2\big(1-\exp[-2a_Tt]\big)/a_T}\bigg].
\label{transition-density}
\end{align}
The representation \eqref{transition-density} follows 
from the property that, given $T(s)=x$, the value of $T(t+s)$ is normally distributed with mean
\begin{equation}
\exp[-a_Tt]x+T_0\big(1-\exp[-a_Tt]\big)
\end{equation}
and variance
\begin{equation}
\frac{\sigma_T^2}{2a_T}\big(1-\exp[-2a_Tt]\big)
\end{equation}
(see \cite[Section 3.3.1]{Glasserman2003}).

\subsection{Cracks}}
 
We consider a band containing straight-line cracks 
perpendicular to the travelling direction. The positions of the cracks in 
the longitudinal direction of the band are described by a counting process
\begin{equation}
N_\xi=\{N_\xi(s),\hspace{0.5em}s\ge 0\}.
\end{equation}
{\color{black}The number of cracks in a band of length $S$ is given by the 
 random variable $N_\xi(S)$.
It is assumed that the process $N_\xi$ is independent of the tension process $T$.

Let $s_i$ denote the distance between the first end of the band and the $i$th crack that 
appears in the draw (see Fig. \ref{initials:fig1}).
In the case of constant tension, we assume that the crack distances 
are strictly positive so that more than one crack does not appear in the 
same longitudinal position of the band simultaneously. In this case, more than one crack may occur in the open draw simultaneously, but the possible interactions of cracks are not considered in this study. In the case of randomly varying tension, we assume that $s_i-s_{i-1}>\ell$. }

\begin{figure}[h!]
\includegraphics[width=0.5\textwidth]{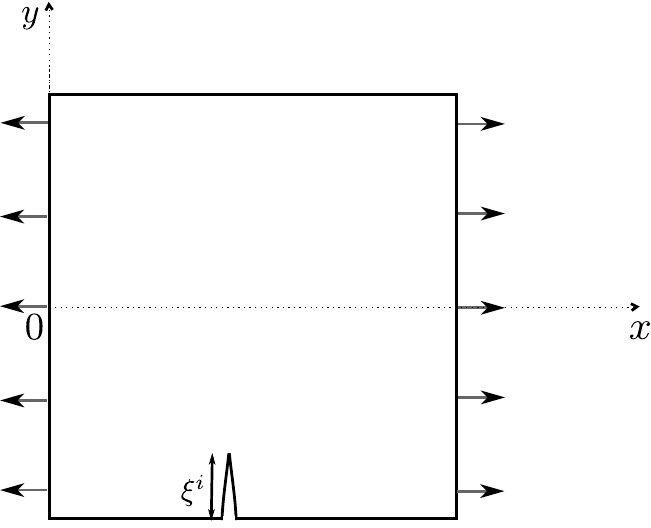}
%
%
\caption{{\color{black}An edge crack on the tensioned plate.}
}
\label{crackfig}       
\end{figure}

In this study, we consider a band containing only {\color{black}through-thickness edge cracks} (see Fig. \ref{crackfig}). The length of the $i$th crack is described by the random variable $\xi^i$. We assume that the random variables $\xi^i$ are independent and identically distributed (i.i.d.), and the common cumulative distribution and probability density functions of the
 crack lengths are denoted by $F_\xi$ and $f_\xi$.  
 The random variables $\xi^i$ are assumed to be independent of the 
 processes $N_\xi$ and $T$. 

{\color{black}Although we consider only sharp edge cracks in this study, the reliability analysis can be generalized for other crack geometries as well by modifying the fracture criterion presented below. For example, instead of describing only the length of a crack as a stochastic quantity, the geometry of the crack can be described by a random vector, the elements of which describe the crack length, the location of the crack in the $y$ direction and the orientation of the crack in the $xy$ plane.

\subsection{Nonfracture criterion}}

To study the fracture of the band, we apply linear elastic fracture mechanics (LEFM), which assumes that 
the inelastic deformation at the crack tip is small compared to the size of the crack. 
{\color{black}Crack loadings in the system are of mode $\MakeUppercase{\romannumeral 1}$ (opening).} 
When a crack $\xi^i$ travels through the open draw, the 
stress intensity factor $K$ related to the crack is a function of the form (see \cite{Fett2008})
\begin{equation}
K(t,\xi^i)=\frac{\alpha(t,\xi^i)\ T(s^i+t)\sqrt{\pi{\color{black}\xi^i}}}{h},\hspace{0.5em}t\in[0,\ell],\label{K}
\end{equation}
where $\alpha$ is a weight function related to the crack geometry.
In this study, we assume that the function $\alpha$ is constant with respect to the location of the crack 
in $x$ direction{\color{black}:}
\begin{equation}
\alpha(t,\xi^i)=\alpha(\xi^i).
\end{equation}
Weight functions for cracks in a rectangular plate under constant tensile loading  are provided,
 for example, in \cite{Perez2004book, Fett2008}. 

The nonfracture criterion for the band when the crack $\xi^i$ travels through the open draw reads as
\begin{equation}
K(t,\xi^i)< K_C\hspace{0.5em}\text{for all}\hspace{0.5em}t\in[0,\ell],\label{constr-for-single-crack}
\end{equation}
where $K_C$ is the fracture toughness of the material. The nonfracture criterion 
\eqref{constr-for-single-crack} is equivalent to
\begin{equation}
T(s^i+t)<B(\xi^i),\hspace{0.5em}t\in[0,\ell],
\end{equation}
with
\begin{equation}
B(\xi^i)=\frac{hK_C}{\alpha(\xi^i)\sqrt{\pi\xi^i}}.
\end{equation}

\begin{figure}[h!]
\centering
\includegraphics[width=0.6\textwidth]{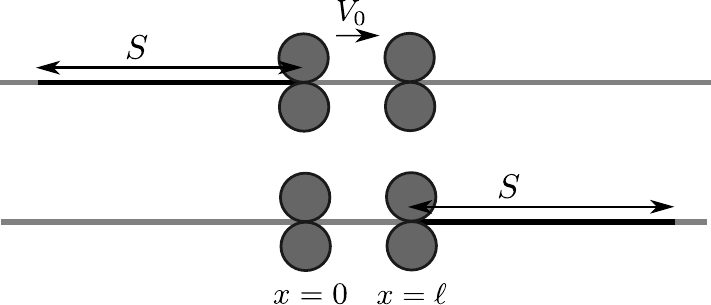}
\caption{The initial and last states of the system.}\label{transition}
\end{figure}

 The performance of the system is considered during the transition of a band of length $S$
through the open draw. In this, 
the initial and last states of the system are regarded as the states at which the first and last ends of the band 
are located at the supports to which the travelling material arrives first and last, respectively (see Fig. \ref{transition}). It is assumed that
    before and after the band the material continues and remains similar. For simplicity, cracks that occur in the open draw in the initial and last states are not considered in terms of fracture.

\section{Reliability in terms of fracture}

In this section, representations for the reliability of the system are derived with different tension models. For a general counting process describing the crack occurrence, the reliability of the system can be obtained by utilizing conditional Monte Carlo sampling. {\color{black}Explicit representations are {\color{black}derived} for  
special cases.}

\subsection{Constant tension}\label{const-tens}

When tension is constant and the possible interactions of the cracks that occur in the open draw simultaneously are not taken into account,
the reliability of the system reads as
\begin{align}
r_1=&\mathbb{P}\big[N_\xi(S)=0\big]\\
&+\mathbb{P}\big[N_\xi(S)\ge 1,\hspace{0.5em}T_0<B(\xi^i)
\hspace{0.5em}\text{for all}\hspace{0.5em}i=1,\dots,N_\xi(S)\big].\label{reliab1}
\end{align}
Since $N_\xi$ 
is independent of the crack lengths, and the lengths are i.i.d., it holds that
\begin{equation}
r_1=\mathbb{P}\big[N_\xi(S)=0\big]+\sum_{j=1}^\infty\mathbb{P}\big[N_\xi(S)=j\big]
\overline{q}^j\label{reliab1-2}
\end{equation}
with
\begin{equation}
\overline{q}=\mathbb{P}\big[T_0<B(\xi^1)\big].\label{const-tens-prob}
\end{equation}
The probability $r_1$ can also be estimated by exploiting the idea of conditional Monte Carlo simulation (see \cite[Section 5.4]{RubinsteinKroesebook2007}). That is, we may estimate
\begin{align}
r_1&\approx\frac{1}{M}\sum_{j=1}^M\chi_{\{k_j=0\}}\label{r1-mc1}\\
&+\frac{1}{M}\sum_{j=1}^M\chi_{\{k_j\neq 0\}}\mathbb{P}
\big[T_0<B(\xi^1),\dots,T_0<B(\xi^{N_\xi(S)})\mid N_\xi(S)=k_j\big],\label{r1-mc2}
\end{align}
where $k_1,\dots,k_M$ is a sample of size $M$ from the distribution of $N_\xi(S)$, and for the conditional probability in \eqref{r1-mc2}, it holds that
\begin{align}
&\mathbb{P}
\big[T_0<B(\xi^1),\dots,T_0<B(\xi^{N_\xi(S)})\mid N_\xi(S)=k_j\big]\\
&=\mathbb{P}
\big[T_0<B(\xi^1),\dots,T_0<B(\xi^{k_j})\big]\\
&=
\overline{q}^{k_j}.
\end{align}

\subsection{Stochastic volatility in tension}\label{stoch-vol}

When there is stochastic volatility in the value of tension, the probability that 
a band of length $S$ travels through the open draw such that a fracture does not 
propagate from any of its cracks is
\begin{align}
r_2=&\mathbb{P}[N_\xi(S)=0]\\
&+\mathbb{P}\big[N_\xi(S)\ge1,\hspace{0.5em}T(s_i+t)<B(\xi^i)\\
&\hspace{3em}
\forall\hspace{0.5em}
t\in[0,\ell]\hspace{0.5em}\forall\hspace{0.5em}i=1,\dots,N_\xi(S)\big].
\end{align}
Similar to Section \ref{const-tens}, 
we may estimate $r_2$ by exploiting conditional Monte Carlo simulation. First, we estimate
\begin{align}
r_2&\approx\frac{1}{M}\sum_{j=1}^M\chi_{\{x^j_1>S\}}+\frac{1}{M}\sum_{j=1}^M\chi_{\{x^j_1\le S\}}
\overline{q}^{j}_{k_j},\label{r2-mc}
\end{align}
where
\begin{align}
\overline{q}_{k_j}^j=\mathbb{P}\big[&T(s_i+t)<B(\xi^i)\hspace{0.5em}\forall\hspace{0.5em}
t\in[0,\ell]\hspace{0.5em}\forall\hspace{0.5em}i=1,\dots,N_\xi(S)\\
&\mid s_1=x_1^j,\dots,s_{k_j}=x_{k_j}^j,s_{k_{j}+1}=x_{k_{j}+1}^j]
\end{align}
and the vectors  $(x_1^j,\dots,x_{k_{j}+1}^j)$, $j=1,\dots,M$ consist of simulated crack distances, satisfying
\begin{equation}
x_1^j+\dots+x_{k_j}^j\le S<x_1^j+\dots+x_{k_{j}+1}^j
.
\end{equation}

The probability $\overline{q}_{k_j}^j$ above simplifies to
\begin{align}
\overline{q}_{k_j}^j=\mathbb{P}\big[&T(x_i^j+t)<B(\xi^i)\hspace{0.5em}\forall\hspace{0.5em}
t\in[0,\ell]\hspace{0.5em}\forall\hspace{0.5em}i=1,\dots,k_j\\
&\mid s_1=x_1^j,\dots,s_{k_j}=x_{k_j}^j,s_{k_{j}+1}=x_{k_{j}+1}^j]\\
=\mathbb{P}\big[&T(x_i^j+t)<B(\xi^i)\hspace{0.5em}\forall\hspace{0.5em}
t\in[0,\ell]\hspace{0.5em}\forall\hspace{0.5em}i=1,\dots,k_j\big].
\end{align}
Since
 $s_j>s_{j-1}+\ell$, we obtain by using the Markov property of $T$ and the independence 
 of $\xi^i$'s that
\begin{align}
\overline{q}_{k_j}^j
&=\mathbb{P}\big[T(x_{k_j}^j+t)<B(\xi^{k_j})\hspace{0.5em}\forall\hspace{0.5em}t\in(0,\ell]\mid T(x_{k_j}^j)<B(\xi^{k_j})]\cdot\label{line3a}\\
&\hspace{1em}\cdot\mathbb{P}\big[T(x_{k_j}^j)<B(\xi^{k_j}),T(x_{i}^j+t)<B(\xi^{i})\label{line1}\\
&\hspace{3em}\forall\hspace{0.5em}t\in[0,\ell]\hspace{0.5em}\forall\hspace{0.5em}i=1,\dots,k_{j}-1\big],\label{line2}
\end{align}
where 
the probability \eqref{line1}--\eqref{line2} is equal to
\begin{equation}
\mathbb{P}\big[T(x_{k_j}^j)<B(\xi^{k_j})\mid T(x_{k_j-1}^{j}+\ell)<B(\xi^{k_j-1})\big]\overline{q}_{k_j-1}^j.
\end{equation}

By the stationarity of $T$ and the assumption that $\xi^i$'s are identically distributed, 
the  probability on the right of \eqref{line3a} simplifies to
\begin{equation}
\frac{q_1}{q_2}
\end{equation}
with
\begin{equation}
q_1=\mathbb{P}\big[T(t)<B(\xi^1)\hspace{0.5em}\forall\hspace{0.5em}t\in[0,\ell]\big]
\end{equation}
and
\begin{equation}
q_2=\mathbb{P}[T(0)<B(\xi^1)].\label{prob2}
\end{equation}
Further, 
we may write
\begin{equation}
q_1
=\int_{\mathbb{R}^+}\mathbb{P}\big[T(t)<B(x)\hspace{0.5em}\forall\hspace{0.5em}t\in[0,\ell]\big]f_\xi(x)dx.
\end{equation}

Let 
\begin{equation}
\tau_{y}^x:=\inf\{s\ge 0:T(s)=x\mid T(0)=y\}\label{hitting-time}
\end{equation}
denote the first passage time (hitting time) of the tension process to the boundary 
$x$ given that the process started at $y$. With this notation we have
\begin{equation}
q_1=\int_{\mathbb{R}^+}\int_{-\infty}^{B(x)}\mathbb{P}[\tau_y^{B(x)}>\ell]f_T(y)f_\xi(x)dydx.\label{prob1}
\end{equation}
The spectral expansion of the survival function of $\tau_y^x$ is given in \cite{Linetsky2004}. According 
to \cite{Linetsky2004}, when $y<x$, it holds that
\begin{equation}
\mathbb{P}[\tau_y^x>s]=\sum_{n=1}^\infty c_ne^{-\lambda_n s},\hspace{0.5em}s>0,\label{series-expr1}
\end{equation}
where $\{\lambda_n\}_{n=1}^\infty$ and $\{c_n\}_{n=1}^\infty$ are obtained as follows: Let
\begin{equation}
\lambda_n=a_T\nu_n,\hspace{0.5em}\overline{x}=-\frac{\sqrt{2a_T}}{\sigma_T}(x-T_0),\hspace{0.5em}\overline{y}=-\frac{\sqrt{2a_T}}{\sigma_T}
(y-T_0).\label{series-expr2}
\end{equation}
The coefficients $\{\nu_n\}_{n=1}^\infty$, $0<\nu_1<\nu_2<...$, $\nu_n\to\infty$ as $n\to\infty$,
 are the positive roots of the equation
\begin{equation}
H_\nu\big(\overline{x}/\sqrt{2}\big)=0,\label{series-expr3}
\end{equation}
where $H_\nu$ is the Hermite function, and the equation is solved with respect to $\nu$.
 The coefficients $\{c_n\}_{n=1}^\infty$ are given by
\begin{equation}
c_n=-\frac{H_{\nu_n}\big(\overline{y}/\sqrt{2}\big)}{\nu_n\frac{\partial}{\partial\nu}
\bigg\{H_\nu\big(\overline{x}/\sqrt{2}\big)\bigg\}\bigg|_{\nu=\nu_n}}.\label{series-expr4}
\end{equation}

Further, we may write
\begin{equation}
\mathbb{P}\big[T(x_{k_j}^j)<B(\xi^{k_j})\mid T(x_{k_j-1}^j+\ell)<B(\xi^{k_j-1})\big]=\frac{q_3^*(x_{k_j-1}^j,x_{k_j}^j)}{q_2}
\end{equation}
with
\begin{align}
q_3^*(x_{k_{j}-1}^j,x_{k_{j}}^j)&=\mathbb{P}\big[T(x_{k_j}^j)<B(\xi^{k_j}),T(x_{k_{j}-1}^j+\ell)<B(\xi^{k_{j}-1})\big]\\
&=\int_{\mathbb{R}^+}\int_{\mathbb{R}^+}\mathbb{P}
\big[T(x_{k_j}^j)<B(x),T(x_{k_{j}-1}^j+\ell)<B(z)\big]\cdot\\
&\hspace{10em}\cdot f_\xi(x)f_\xi(z)dxdz.
\end{align}
Moreover, we have
\begin{align}
&\mathbb{P}
\big[T(x_{k_j}^j)<B(x),T(x_{k_j-1}^j+\ell)<B(z)\big]\\
&=\int_{-\infty}^{B(x)}\int_{-\infty}^{B(z)}
p(x_{k_j}^j-x_{k_j-1}^j-\ell,u,v)f_T(u)dudv,
\end{align}
where $p$ is the transition probability density defined in \eqref{transition-density}.
Thus,
\begin{equation}
q_3^*(x_{k_{j}-1}^j,x_{k_{j}}^j)=q_3(x_{k_j}^j-x_{k_{j}-1}^j)
\end{equation}
with
\begin{align}
q_3(s)
&=\int_{\mathbb{R}^+}\int_{\mathbb{R}^+}
\int_{-\infty}^{B(x)}\int_{-\infty}^{B(z)}p(s-\ell,u,v)\cdot\nonumber\\
&\hspace{13em}\cdot f_T(u)f_\xi(x)f_\xi(z)dudvdxdz.\label{prob3a}
\end{align}
Finally, we notice that \eqref{prob3a} is equivalent to
\begin{align}
q_3(s)&=\int_{\mathbb{R}^+}\int_{\mathbb{R}^+}
\int_{-\infty}^{B(z)}F_{Gauss}\big(\mu_{Gauss}(u,s),\sigma_{Gauss}(s),B(x)\big)\cdot\nonumber\\
&\hspace{13em}\cdot f_T(u)f_\xi(x)f_\xi(z)dudxdz,\label{prob3}
\end{align}
where 
$F_{Gauss}\big(\mu_{Gauss}(u,s),\sigma_{Gauss}(s),x\big)$ denotes the cumulative 
distribution function of the normal random 
variable with
mean 
\begin{equation}
\mu_{Gauss}(u,s)=T_0+(u-T_0)e^{-a_T(s-\ell)}
\end{equation}
and standard deviation
\begin{equation}
\sigma_{Gauss}(s)=\sigma_T\sqrt{\frac{1-e^{-2a_T(s-\ell)}}{2a_T}}
\end{equation}
at point $x$.

By the same reasoning as above, it holds for all $i=2,\dots,k_j-1$ that
\begin{equation}
\overline{q}_{i}^j=\frac{q_1q_3(x_{i}^j-x_{i-1}^j)}{q_2^2}\overline{q}_{i-1}^j.
\end{equation}
In addition,
\begin{equation}
\overline{q}_1^j=q_1.
\end{equation}
Accordingly, it holds that
\begin{equation}
\overline{q}_{k_j}^j=q_1\bigg(\frac{q_1}{q_2^2}\bigg)^{k_j-1}\prod_{i=2}^{k_j}q_3(x_{i}^j-x_{{i-1}}^j).\label{qj}
\end{equation}

\subsection{Examples}\label{r1-r2-examples}

As examples, we consider cases in which cracks occur in the band according to renewal processes. For such a process, 
the distances between succeeding cracks are independent and identically 
distributed. 

As an example, we consider the reliability of the system when the tension is constant, and cracks occur in the band according to a homogeneous Poisson process with intensity $\lambda_\xi$. In this case, the expected distance of two succeeding cracks is $1/\lambda_\xi$. The representation \eqref{reliab1-2} is equivalent to 
\begin{equation}
r_1=e^{-\lambda_\xi S}\sum_{j=0}^\infty \frac{(\lambda_\xi S)^j}{j!}\overline{q}^j.\label{constr4}
\end{equation}
Noticing that the series in \eqref{constr4} is the Maclaurin series of the exponential function at point
$\lambda_\xi S \overline{q}$,
the formula \eqref{constr4} can be written as
\begin{equation}
r_1=\exp(\lambda_\xi S(\overline{q}-1)).\label{ex}
\end{equation}

Another example is given by the case in which defects occur (almost) periodically in some part of the band. 
When the possible crack locations in the longitudinal direction of the band are
\begin{equation}
iL,\hspace{0.5em}i=1,\dots,\lfloor \overline{S}/L,\rfloor,\hspace{0.5em} \overline{S}\le S,\hspace{0.5em}L>\ell,\label{r1-binom-model}
\end{equation}
 and a crack occurs in location $iL$ with probability $p_s$ independently of other cracks, 
 the random variable $N_\xi(S)$ follows the binomial distribution with number of trials 
 $\lfloor \overline{S}/L\rfloor$ and a succes probability $p_s$ in each trial. 
 The reliability of the system with constant tension reads as
\begin{align}
r_1&=(1-p_s)^{\lfloor \overline{S}/L\rfloor}+\sum_{j=1}^{\lfloor \overline{S}/L\rfloor}\binom {\lfloor \overline{S}/L\rfloor}{j}(p_s)^j(1-p_s)^{\lfloor \overline{S}/L\rfloor-j}
\overline{q}^j\\
&=(1+p_s(\overline{q}-1))^{\lfloor \overline{S}/L\rfloor}.\label{r1-bin}
\end{align}
To simulate the reliability with tension variations, we notice that
 \begin{equation}
s_i-s_{i-1}=LX,
 \end{equation}
 where $X$ follows the geometric distribution with the success probability $p_s$ and the support $\{1,2,\dots\}$. The expected distance between cracks is 
 \begin{equation}
 \mathbb{E}[s_i-s_{i-1}]=\frac{L}{p_s}.
 \end{equation}

When the distance between two succeeding cracks is a constant $L$, the reliability of the system is
\begin{equation}
r_1=\overline{q}^{\lfloor S/L\rfloor},\hspace{0.5em}L>0\label{r1-const}
\end{equation}
when tension is constant, and 
\begin{equation}
r_2=q_1\bigg(\frac{q_1q_3(L)}{q_2^2}\bigg)^{\lfloor S/L\rfloor-1},\hspace{0.5em}L>\ell\label{r2-det}
\end{equation}
when there is stochastic volatility in tension.

In the numerical examples, we also consider the case in which the distances between two succeeding cracks obey the $3$-parameter lognormal 
distribution with the support $(\ell,\infty)$. Denoting the common probability density function 
of the crack distances by $f_{s_\xi}$, we have 
\begin{equation}
f_{s_\xi}(x)=\frac{1}{\sigma_{s}(x-\ell)\sqrt{2\pi}}\exp\bigg[-\frac{\big(\ln (x-\ell)-\mu_{s}\big)^2}{2\sigma_{s}^2}\bigg],\hspace{0.5em}
x>\ell,
\end{equation}
with shape $\sigma_s>0$ and log-scale $\mu_s\in\mathbb{R}$. 
With the $3$-parameter lognormal distribution, the
 expected distance between cracks is
 \begin{equation}
 \mathbb{E}[s_i-s_{i-1}]=\ell+e^{\mu_s+\sigma_s^2/2},
 \end{equation}
 and the variance of the distance is
 \begin{equation}
 \text{Var}[s_i-s_{i-1}]=e^{2\mu_s+\sigma_s^2}\big(e^{\sigma_s^2}-1\big).
 \end{equation}

  For all the models, we assume that the distance between the first crack and the first end of the band has the same distribution, or is the same, as the distance between the two succeeding cracks.

The reliability decreases when the tension increases, and thus
we may seek the critical value of tension such that the safe transition of a band of length $S$ through the open draw is guaranteed at a given level. In the case of constant tension, the problem reads as
\begin{align}
&\max T_0\hspace{0.5em}\text{such that}\label{problem}\\
&\hspace{0.3em}r_1\ge q,\label{constr}
\end{align}
where $q\in(0,1)$ is the required reliability level.  
Let the crack length $\xi^i$ obey a continuous distribution with the support $\mathbb{R}^+$. 
Assuming that the function $g(x)=\alpha(x)\sqrt{x}$ is strictly increasing, it holds that
\begin{equation}
\overline{q}=F_\xi\bigg(g^{-1}\bigg(\frac{hK_C}{T_0\sqrt{\pi}}\bigg)\bigg),\label{p1p2}
\end{equation}
where $g^{-1}$ denotes the inverse function of $g$.
When $N_\xi$ is a homogeneous Poisson process, the solution 
of \eqref{problem}--\eqref{constr} is
\begin{equation}
T_0^{cr}=\frac{hK_C}{\sqrt{\pi}}\bigg( g\bigg(F^{-1}_\xi\bigg(\frac{\log(q)}{\lambda_\xi S}+1\bigg)\bigg)\bigg)^{-1},\label{cr-tension}
\end{equation}
where $F^{-1}_\xi$ denotes the inverse function of $F_\xi$. 
When $N_\xi(S)$ obeys the binomial distribution, the critical velocity is
\begin{equation}
T_0^{cr}=\frac{hK_C}{\sqrt{\pi}}\bigg(g\bigg(F^{-1}_\xi\bigg(\frac{q^{1/\lfloor \overline{S}/L\rfloor}-1}{p_s}+1\bigg)\bigg)\bigg)^{-1}\label{cr-tension-binom}
\end{equation}
when tension is constant.

\section{Numerical examples and discussion}

The reliability of the system was computed with different models for tension and crack occurrence. The values of the {\color{black}material and machine} parameters used in the examples are typical
of dry paper (newsprint) and printing presses. 

\subsection{Numerical solution process and error approximation}

The computations were carried out with Mathematica, in which a built-in function for 
the Hermite function appearing in the construction of the series \eqref{series-expr1} is available.
The roots $\{\lambda_n\}_{n=1}^\infty$ of the Hermite function were 
sought by combining the plain bisection method and Mathematica's \emph{FindRoot} function using the Brent method. Intervals that bracket 
the roots were found by starting from the preceeding root, or zero in the case of the first root, and 
computing the values of 
the Hermite function in \eqref{series-expr3} step by step until its sign had changed with such a small
step size that no roots were skipped. 
 The series \eqref{series-expr1} was 
truncated after the $k$th term that was the first to satisfy
\begin{equation}
c_ne^{-\lambda_n S}\le 10^{-16}.
\end{equation}
In computing the coefficients $\{c_n\}_{n=1}^\infty$, a readily available 
numerical derivation function in Mathematica was utilized.

Mathematica's \emph{NIntegrate} function was used to compute
 estimate for the integrals {\color{black}$\overline{q}$ and $q_2$. The probabilities $q_1$ and $q_3$} were estimated by Monte Carlo simulation. In the computations, the errors of the Monte Carlo estimates were approximated by the standard error (see Section 1.1.1. in \cite{Glasserman2003}). 
 
The error in \eqref{qj} that originates from the
error
of the integrals $q_1$, $q_2$ and $q_3(x_i^j-x_{i-1}^j)$, $i=2,\dots,k_j$ was approximated by its total differential. That is, when the computed estimates of these integrals differ from the exact values by small quantities $dq_i$, the corresponding error in \eqref{qj} can be approximated by
\begin{equation}
d \overline{q}_{k_j}^j=\frac{\partial \overline{q}_{k_j}^j}{\partial q_1}d q_1+\frac{\partial \overline{q}_{k_j}^j}{\partial q_2}d q_2 +\sum_{i=2}^{k_j}\frac{\partial \overline{q}_{k_j}^j}{\partial \big(q_3(x_{i}^j-x_{{i-1}}^j)\big)}d q_3(x_{i}^j-x_{{i-1}}^j).
\end{equation}
It holds that
\begin{align}
d \overline{q}_{k_j}^j&\le k_j\frac{\overline{q}_{k_j}^j}{q_1}|d q_1|+2|1-k_j|\frac{\overline{q}_{k_j}^j}{q_2}|d q_2|\\
&\hspace{5em} +\sum_{i=2}^{k_j}\frac{\overline{q}_{k_j}^j}{q_3(x_{i}^j-x_{{i-1}}^j)}|d q_3(x_{i}^j-x_{{i-1}}^j)|\\
& \le k_j|d q_1|+2|1-k_j||d q_2| +(k_j-1)\max_{i=2,\dots,k_j}|d q_3(x_{i}^j-x_{{i-1}}^j)|
\end{align}
since the terms $\overline{q}_{k_j}^j/q_1$, $\overline{q}_{k_j}^j/q_2$ and $\overline{q}_{k_j}^j/q_3(x_{i}^j-x_{{i-1}}^j)$, $i=2,\dots,k_j$ can be regarded as conditional probabilities and thus are not more than one. Consequently, when
\begin{equation}
|d q_1|+2|dq_2|+\max_{i=2,\dots,k_j}|d q_3(x_{i}^j-x_{{i-1}}^j)|\le\epsilon,
\end{equation}
 we may approximate
\begin{equation}
d\bigg(\frac{1}{M}\sum_{j=1}^M\chi_{\{x_1^j\le S\}}\overline{q}_{k_j}^j\bigg)\le \epsilon\max_{j=1,\dots,M}k_j.\label{error-upper-bound}
\end{equation}
Similarly, if the error in $\overline{q}$ is bounded above by $\epsilon$, the same upper bound as in \eqref{error-upper-bound} is obtained for the  error in \eqref{r1-mc2}. For the explicit formulae \eqref{ex}, \eqref{r1-bin}, \eqref{r1-const} and \eqref{r2-det}, the error can be approximated in a similar manner. 

\subsection{Examples for printing presses}

The values of the {\color{black}machine and material} parameters used in computing the examples of this section are typical of those of printing  presses and 
dry paper (newsprint). {\color{black}Values of the deterministic parameters} are listed in Table
\ref{tab:parameters}.
  The strain energy release rate $G_{\mathrm{C}}$ was obtained from the 
  results in \cite{SethPage1974}, and the fracture toughness was set to 
  \begin{equation}
  K_C=\sqrt{G_CE}.
  \end{equation}
  The band length was given the value $S=350$ (km). Uesaka \cite{Uesaka2013} approximates that an average 
 distance between web breaks in a printing press is $350$ km. 
\begin{table}
\centering
\begin{tabular}{| l | l | }
\hline
$\ell$ & $1\hspace{0.5em}\text{(m)}$ \\
$b$ & $0.6\hspace{0.5em}\text{(m)}$ \\ 
  $h$ &  $8\cdot10^{-5}\hspace{0.5em}\text{(m)}$ \\
  $E$ & $4\hspace{0.5em}\text{(GPa)}$  \\
  $G_{\mathrm{C}}$ & $6500\hspace{0.5em}\text{(J}/\text{m}^2)$\\
  \hline  
\end{tabular}
\caption{Deterministic parameter values.}
\label{tab:parameters}
\end{table}

 When the values of $c_T$ and $a_T$ are set, the volatility 
parameter $\sigma_T$ is obtained from Equation \eqref{c-T}. In the computations, it was set $a_T=1$, and the reliability of the system was studied with $T_0=200$, $350$, $500$ (N/m) and {\color{black}$c_T=0.05$, $0.1$}. For the tension values usually applied in printing presses, see the measurements in \cite{Uesaka2004,linnabetter}.

\begin{figure}[h!]
\includegraphics[width=0.45\textwidth]{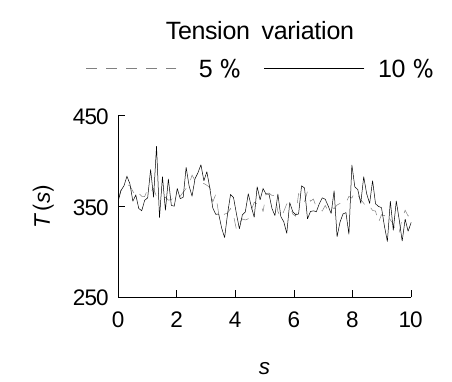}
\caption{
{\color{black}A sample path of the tension process with different values of $c_T$ with    $T_0=350$ (N/m) and $a_T=1$.}}
\label{tensionprocess}      
\end{figure}

Single simulated sample paths of the tension process are shown in 
Figure \ref{tensionprocess} 
with different values of $c_T$ with $T_0=350$ (N/m).
Discretization of the Ornstein-Uhlenbeck process is represented, for example, in \cite[Section 3.3.1]{Glasserman2003}.
In the figure, $100$ discretization points were used for the considered
interval. {\color{black}For comparison, see \cite[Figure 2]{roisum1990}.}

The weight function $\alpha$ that appears in the stress intensity factor \eqref{K} was approximated 
from the results in {\color{black}\cite[Section
C8.1]{Fett2008}}. That is, it was set to
\begin{equation}
\alpha(\xi^i)={\color{black}\frac{F'\big(\xi^i/(2b)\big)}{\big(1-\xi^i/(2b)\big)^{3/2}}},
\end{equation}
where the function $F'$ was interpolated by using Mathematica's \emph{Interpolation} function from the values 
 in {\color{black}\cite[Table C8.1]{Fett2008}}.

{\color{black}The reliability of the system was studied with Weibull distributed crack lengths. In \cite{swineharBroekt1996}, the distribution of holes in a paper web was represented by a Weibull distribution. With this crack length model, the distribution function of the crack length is \cite[Section 4]{siegrist2003}
\begin{equation}
F_\xi(x)=1-e^{-(x/\lambda_\xi)^{k_\xi}},\hspace{0.5em}x\ge 0,
\end{equation}
where $\lambda_\xi>0$ and $k_\xi>0$ are the scale and shape parameters of the distribution. The mean and the variance of the crack length are  \cite[Section 4]{siegrist2003}
\begin{equation}
\mathbb{E}[\xi^i]=\lambda_\xi\Gamma(1+1/k_\xi)
\end{equation}
and 
\begin{equation}
\text{Var}[\xi^i]=\lambda_\xi^2\bigg[\Gamma\bigg(1+\frac{2}{k_\xi}\bigg)-\bigg(\Gamma\bigg(1+\frac{1}{k_\xi}\bigg)\bigg)^2\bigg].
\end{equation}

The examples were computed with $k_\xi=0.8$ which is comparable to the shape parameter of the hole size distribution in \cite{swineharBroekt1996}. With this, independent of $\lambda_\xi$, the coefficient of variation (the standard deviation divided by the mean) of the crack length is $1.26$. The reliability of the system was studied with different values of the expected crack length.}

The reliability of the system with constant tension 
was studied with the Poisson, binomial, lognormal and deterministic crack occurrence models introduced in Section \ref{r1-r2-examples}. 
 The lognormal model was examined with two different values for the coefficient of variation of the crack distances: one and ten. With the binomial model, it was set $L=2$ and $p_s=0.9$.
The reliability of the system with tension variations was considered with the binomial and deterministic crack occurrence models. 

In general, the sample size in computing {\color{black} $q_1$ and $q_3$} and the accuracy goal for $\overline{q}$ and $q_2$ were chosen such that the estimated errors
 in $r_1$ and $r_2$ were approximately $0.01$ at maximum. {\color{black}However, for $T_0=350$, $500$ (N/m), the maximum error of $0.035$ was allowed in computing $r_2$ for the binomial crack occurrence model. In addition, for $T_0=350$ (N/m), the maximum error $0.025$ was allowed in computing $r_2$ for the deterministic crack occurrence model with the smallest crack distance $100$ (m).} In simulating the reliability with constant tension and the lognormal crack occurrence model, a sample size of $M=100$ in \eqref{r1-mc1}--\eqref{r1-mc2} was used. With this sample size, the standard errors of the estimates for $r_1$ were approximately {\color{black} $5\cdot 10^{-6}$} at maximum. With the binomial crack occurrence model and tension variations, the sample size $M=100$ in \eqref{r2-mc} was used. This produced standard errors for the estimates less than {\color{black}$2\cdot 10^{-4}$}.

\begin{figure}[h!]
\includegraphics[width=\textwidth]{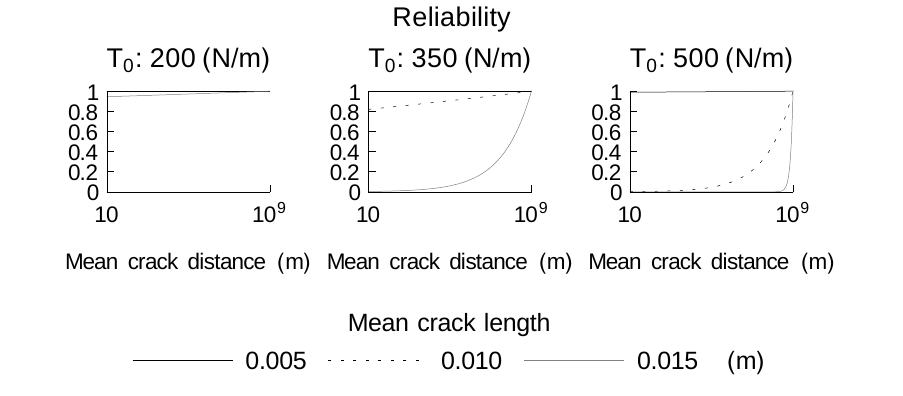}
\caption{
{\color{black}Reliability of the system with Poisson model for crack ocurrence. Constant tension.}}
\label{r1poisson}      
\end{figure}

Figure \ref{r1poisson} shows the reliability of the 
 system with constant tension when cracks occur according to a Poisson process. The impact of the mean crack length on the reliability of the system increased when the tension increased. For the studied values of tension, the change was notable: For example, with $T_0=200$ (N/m) and {\color{black}$\mathbb{E}[s_i-s_{i-1}]=10^8$ (m)}, the reliability of the system decreased from {\color{black}$1.0$ to $0.95$} when the mean crack length increased from {\color{black}$0.005$ (m) to $0.015$ (m)}. With $T_0=350$ (N/m), the corresponding reliabilities were {\color{black}$1.0$ and $0.01$}. {\color{black}Also}, the mean distance between cracks was a considerable factor in terms of the system reliability: For example, with $T_0=500$ (N/m) and $\mathbb{E}[\xi^i]=0.01$ (m), the reliability  was only {\color{black}$0.05$ with $\mathbb{E}[s_i-s_{i-1}]=5\cdot 10^8$ (m)} but increased to {\color{black}$1.0$} when the distance increased to {\color{black}$10^9$ (m). Moreover, it is seen that when the mean crack length is only $0.005$ (m), cracks do not affect the reliability of the system, even when the mean distance between cracks is small or tension is high. On the other hand, when the mean crack length is larger and tension is high, cracks may affect the reliability of the system, unless the mean distance between cracks is extremely large.}

With the studied parameter values, no remarkable difference in the reliability of the system with constant tension was found between the lognormal and deterministic crack occurrence models when the average distance between cracks in the lognormal model was set to be equal to the distance between cracks in the deterministic model. Naturally, the difference between the deterministic and lognormal models was larger with the higher coefficient of variation of the crack distances. The maximum difference was approximately $0.03$.

\begin{figure}[h!]
\centering
\includegraphics[width=\textwidth]{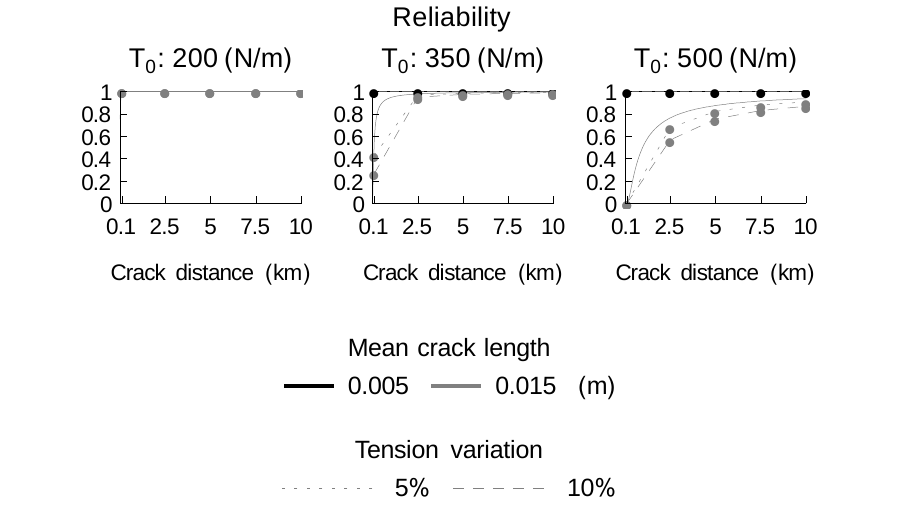}
\caption{
{\color{black}Effect of stochastic volatility in tension on reliability. Deterministic model for crack occurrence.}}
\label{stoch-vol-effect-determ}      
\end{figure}

{\color{black}In Figure \ref{stoch-vol-effect-determ}, we see the reliability of the system with the deterministic model for crack occurrence. For the studied crack distances, cracks of mean length $0.005$ (m) did not affect the reliability of the system, even with high average tension and remarkable tension fluctuations. The results suggest that larger cracks ($\mathbb{E}[\xi^i]=0.015$ (m)) may have a greater impact on the system reliability, and the effect of cracks increased significantly when the set tension increased. With $\mathbb{E}[\xi^i]=0.015$ (m) and $T_0=200$ (N/m), the probability of fracture was zero for all studied crack distances but, e.g., when the crack distance was $5$ (km), the reliability $r_1$ decreased to $0.87$ when $T_0$ increased to $500$ (N/m). 
As with the Poisson model, it was seen that the distance between cracks had a considerable impact on the reliability. E.g., with $T_0=500$ (N/m) and $\mathbb{E}[\xi^i]=0.015$ (m), the reliability $r_1$ increased from $0.76$ to $0.91$, when the crack distance increased from $2.5$ to $7.5$ (km). Moreover, the results suggest that tension fluctuations may significantly affect the system reliability. In this, the set tension played an important role. E.g., when $T_0=350$ (N/m), the crack distance was $5$ (km) and $\mathbb{E}[\xi^i]=0.015$ (m), the reliability of the system was close to one ($0.97$) even with $c_T=0.1$. With $T_0=500$ (N/m), the reliability of the system decreased from $0.87$ to $0.75$, when tension fluctuations ($c_T=0.1$) were introduced in the system.}

\begin{figure}[h!]
\centering
\includegraphics[width=\textwidth]{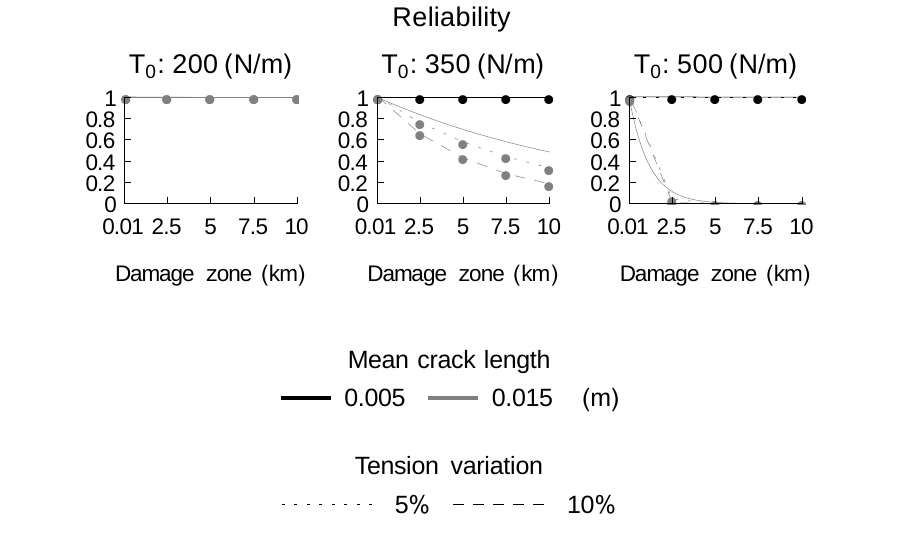}
\caption{
{\color{black}Effect of stochastic volatility in tension on reliability. Binomial model for crack occurrence.}}
\label{stoch-vol-effect-binom}      
\end{figure}

{\color{black}Figure \ref{stoch-vol-effect-binom} shows the reliability of the system with the binomial crack occurrence model. As with the deterministic crack occurrence model, cracks of mean crack length $0.005$ (m) did not affect the system reliability even with high average tension and tension fluctuations. Cracks with larger mean length may affect the system reliability, at least if the tension is not low. For the studied parameter values, the effect of cracks increased significantly when the set tension increased. With $\mathbb{E}[\xi^i]=0.015$ (m) and $T_0=200$ (N/m), the reliability of the system was one in the studied range of damage zone length. With the damage zone length $5$ (km), the reliability $r_1$ decreased to $0.70$ when $T_0$ increased to $350$ (N/m). 
Also, the reliability of the system depended remarkably on the length of the damage zone. E.g., with $T_0=350$ (N/m) and $\mathbb{E}[\xi^i]=0.015$ (m), the reliability $r_1$ decreased from $0.84$ to $0.58$, when the damage zone length increased from $2.5$ to $7.5$ (km). Again, it was seen that tension fluctuations may significantly affect the system reliability. E.g., when $T_0=350$ (N/m), the damage zone length was $2.5$ (km) and $\mathbb{E}[\xi^i]=0.015$ (m), the reliability of the system with constant tension was $0.84$ but with $c_T=0.1$ the reliability was only $0.66$.}

\begin{figure}[h!]
\centering
\includegraphics[width=\textwidth]{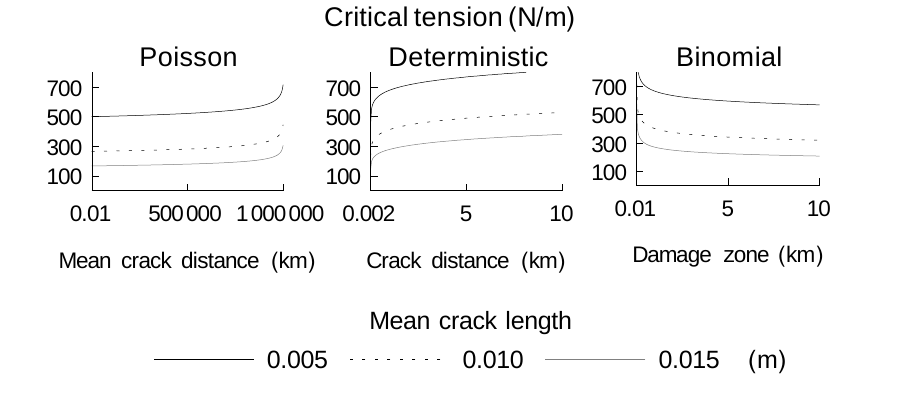}
\caption{
{\color{black}Critical tension with $q=0.99$ when tension is constant.}}
\label{crit-tensions}      
\end{figure}

Figure \eqref{crit-tensions} shows the critical tension for the system with constant tension with the  Poisson, deterministic and binomial models for crack occurrence. In the computations, the required reliability of the system was set to 
 $q=0.99$. {\color{black}To compare, the nominal level of tension in printing presses is $[200,500]$ (N/m) (see \cite{Uesaka2004}). When the mean crack length was $0.015$ (m), the critical tensions were close to the lower bound of the nominal tension. With the average crack length $0.005$ (m), the critical tension can be higher than what is typically applied in printing presses.}%

The computed examples suggest that the set tension has a significant impact on the reliability of the system. When the set tension increases, the impact of {\color{black} cracks becomes more pronounced.} In addition, the impact of tension variations increase remarkably when the set tension increases. {\color{black} With high average tension, tension fluctuations may significantly affect the system reliability.} The results also show that crack frequency is a significant factor in terms of fracture. 

\subsection{Discussion}

{\color{black}In this paper, the reliability of a system with moving cracked material was studied in terms of fracture. Numerical examples were computed with material and machine parameters typical of newsprint and printing presses. However, it should be noted that the numerical results obtained in this study are mainly qualitative, and more rigorous conclusions require data of defects and tension from a real printing press. Such data can be obtained by automated inspection systems developed for quality control \cite{JiangGao2012} and 
devices designed for tension profile measuring \cite{ParolaEtAl2000}. 

In this study, tension fluctuations were described by the stationary Ornstein-Uhlenbeck process.  For such process, a known explicit result for the distribution of the first passage time to a counstant boundary exists and could be exploited in computing. 
However, the results generalize for other stationary and Markov processes at least via simulation of the first passage time distribution.

When the numerical results are considered, it should also be kept in mind that the model lacks some features typical of a moving paper web in a printing press, which may have an 
impact on the results: The study assumed the profile of tension 
to be homogeneous, although in printing presses, the
measured tension varies in the cross-direction (see the measurements in \cite{linnabetter, Uesaka2004}). The results were obtained with the elastic material model, although the paper material is known to have orthotropic characteristics. The study considered the reliability of the system in terms of fracture when the material travels between the supports, but the effect of the rollers was not included in the model.


The present paper extends previous studies of break rate models by modelling tension fluctuations and crack occurrence by a continuous time stochastic process and a stochastic counting process, respectively. The numerical results suggest that tension variations may have a significant impact on the reliability of the system. Thus, including tension fluctuations in the break rate model is essential. The results also show that the fracture probability highly depends on the crack frequency. Thus, upper estimates of the break rate obtained by assuming that a crack exists, e.g., in every roll may lead to overconservative set tension.}

\section{Conclusions}

In this paper, the reliability of processes with moving elastic and isotropic
 material containing initial cracks 
was studied in terms of fracture. The
  material was modelled as a moving plate subjected to homogeneous tension acting in the travelling direction. The 
 reliability of the 
 system was considered in two cases: i)
the tension is constant with respect to time, and ii) the tension varies temporally
according to an Ornstein-Uhlenbeck process. 

The cracks were assumed to occur in the travelling direction according
to a stochastic counting process. {\color{black}Edge cracks perpendicular to the travelling direction were considered.} The lengths of the cracks 
were modelled by i.i.d. random variables. 

For a general counting process describing crack occurrence, a representation for the reliability of the system was derived that exploits conditional Monte Carlo simulation. Explicit formulae were obtained for special cases. In the case of temporally varying tension, considering the fracture probability led to a first passage time problem. Solving this, a known result for the first passage time of an Ornstein-Uhlenbeck process to a constant boundary was utilized.

Numerical examples were provided for parameter values typical of printing presses and paper material. It was seen that the effect of crack length distribution on reliability increased significantly when the set tension increased. The set tension had a remarkable impact on how tension dispersion affected the reliability of the system. Also, crack frequency was an important factor in terms of fracture.

\section{Acknowledgments}

This research was financially supported by the KAUTE Foundation and the Doctoral Program in Computing and Mathematical Sciences
(COMAS).

\section{References}
\bibliographystyle{plain}
\bibliography{tirronen_references}

\end{document}